\documentclass[letterpaper]{jpconf}
\usepackage{graphicx,amsmath}

\newcommand{\bea}{\begin{eqnarray}}
\newcommand{\eea}{\end{eqnarray}}
\newcommand{\be}{\begin{equation}}
\newcommand{\ee}{\end{equation}}
\newcommand{\bes}{\begin{subequations}}
\newcommand{\ees}{\end{subequations}}
\def\lag{\langle}
\def\rag{\rangle}

\def\nbox#1#2{\vcenter{\hrule \hbox{\vrule height#2in
\kern#1in \vrule} \hrule}}
\def\sq{\,\raise.5pt\hbox{$\nbox{.09}{.09}$}\,}
\def\sqb{\,\raise.5pt\hbox{$\overline{\nbox{.09}{.09}}$}\,}

\begin{document}
\pagestyle{plain}
\vspace{-5cm} \hspace{12cm} LANL-UR-11-10829\\

\title{New Horizons in Gravity:\\ Dark Energy and Condensate Stars}

\author{Emil Mottola}

\address{Theoretical Div., Los Alamos National Laboratory, Los Alamos, NM 87545 USA}

\ead{emil@lanl.gov}

\begin{abstract}

Black holes are an apparently unavoidable prediction of classical General Relativity, at 
least if matter obeys the strong energy condition $\rho + 3p \ge 0$. However quantum vacuum 
fluctuations generally violate this condition, as does the eq. of state of cosmological dark energy 
$\rho = - p > 0$. When quantum effects are considered, black holes lead to a number of 
thermodynamic paradoxes associated with the Hawking temperature and assumption of 
black hole entropy, which are briefly reviewed. It is argued that the largest quantum effects arise 
from the conformal scalar degrees of freedom generated by the trace anomaly of the 
stress-energy tensor in curved space. At event horizons these can have macroscopically 
large backreaction effects on the geometry, potentially removing the classical event horizon 
of black hole and cosmological spacetimes, replacing them with a quantum phase boundary
layer, where the effective value of the gravitational vacuum energy density can change. In the 
effective theory including the quantum effects of the anomaly, the cosmological term 
$\Lambda$ becomes a dynamical condensate, whose value depends upon boundary conditions 
at the horizon. By taking a positive value in the interior of a fully collapsed star, the effective
cosmological term removes any singularity, replacing it with a smooth dark energy de Sitter 
interior. The resulting gravitational vacuum condensate star (or {\it gravastar}) configuration 
resolves all black hole paradoxes, and provides a testable alternative to black holes as the 
final quantum mechanical end state of complete gravitational collapse. The observed 
$\Lambda_{eff}$ dark energy of our universe likewise may be a macroscopic finite size 
effect whose value depends not on Planck scale or other microphysics but on the 
cosmological Hubble horizon scale itself.

\end{abstract}

\section{Introduction: Classical Black Holes}
\vspace{3mm}

Just a year after the publication of the field equations of General Relativity 
(GR), K. Schwarzschild found a simple, static, spherically symmetric solution 
of those equations, with the line element \cite{Schw},
\be
ds^2 = -f(r)\, d\tau^2 + \frac{dr^2}{h(r)} + r^2
\left( d\theta^2 + \sin^2\theta\,d\phi^2\right)\,,
\label{sphsta}
\ee
where in this case
\be
f(r) = h(r) = 1 - \frac{r_{\!_M}}{r} = 1 - \frac{2GM}{rc^2} \,.
\label{Sch}
\ee
This Schwarzschild solution to the vacuum Einstein's equations, 
with vanishing Ricci tensor $R^a_{\ b} = 0$ and stress tensor $T^a_{\ b} = 0$
for all $r > 0$ describes an isolated, non-rotating object of total mass $M$.

If the Schwarzschild solution (\ref{sphsta})-(\ref{Sch}) is taken seriously for 
$r< r_{\!_M} = 2GM/rc^2$, the singularity at $r=0$ is present in Einstein's theory 
for any mass $M>0$, including the macroscopic mass of a collapsed star with 
the mass of the sun, $M_{\odot} \simeq 2 \times 10^{33}$ gm. or even that  of 
supermassive objects with masses $10^6$ to $10^9M_{\odot}$. The collapse 
of such enormous quantities of matter to a single mathematical point at $r=0$ 
certainly presents  a challenge to the imagination, and one that it seems Einstein 
himself sought arguments to avoid \cite{Ein}. The situation is scarcely more acceptable 
if the singularity is removed only by the intervention of quantum effects at
the extremely tiny Planck length $L_{Pl} = (G\hbar/c^3)^{\frac{1}{2}} =
1.616 \times 10^{-33}$ cm.

On the other hand, unlike the central singularity of the Schwarzschild metric at $r=0$, and 
despite the coordinate singularity at $r=r_{\!_M}$, local scalar invariant quantities that can be 
constructed from the contractions of the Riemann curvature tensor remain finite at 
$r = r_{\!_M}$. For example the fully contracted quadratic Riemann invariant 
\be
R^{abcd}R_{abcd} = \frac{12\,r_{\!_M}^2}{r^6}\,,
\label{RieS}
\ee
which diverges at $r=0$ remains finite at $r= r_{\!_M}$. Also, although the time for an infalling 
particle to reach the horizon is infinite for any observer remaining fixed outside the horizon, 
the {\it proper time} measured by the particle itself during its infall remains finite as 
$r \rightarrow r_{\!_M}$ \cite{MTW,WeinGR}.  Since the line element (\ref{sphsta}) is non-singular 
also for $0 < r < r_{\!_M}$, the most straightforward possibility would seem to be to assume that this 
non-singular vacuum interior can be matched smoothly to the non-singular exterior Schwarzschild 
solution. This matching was achieved by the coordinate transformations and analytic 
continuation of the Schwarzschild solution found by Kruskal and Szekeres \cite{KrusSz}. 
In the these new $(T,X)$ coordinates, the Schwarzschild line element (\ref{sphsta})-(\ref{Sch})
becomes
\be
ds^2 = \frac{4 r_{\!_M}^3}{r}\, e^{r/2r_{\!_M}}\, (-dT^2 + dX^2) 
+ r^2 \left( d\theta^2 + \sin^2\theta\,d\phi^2\right)\,,
\label{SKSz}
\ee
where $r$ is to be regarded as the function of the Kruskal-Szekeres $(T,X)$ defined implicitly by 
\be
\left(\frac{r}{r_{\!_M}} - 1\right)\, e^{r/r_{\!_M}}= X^2 - T^2\,.
\label{rimp}
\ee
Thus the (singular) transformation to Kruskal-Szekeres coordinates $(T,X)$ has
removed the coordinate singularity of the Schwarzschild line element (\ref{sphsta})-(\ref{Sch}).

An important point which is often left unstated is that this mathematical procedure of analytic 
continuation through the null hypersurface of an event horizon at $r=r_{\!M}$ actually involves 
a {\it physical assumption}, namely that the stress-energy tensor $T^a_{\ b}$ is vanishing there. 
Even in the purely classical theory of General Relativity, the hyperbolic character of 
Einstein's equations allows generically for stress-energy sources and hence metric 
discontinuities on the horizon which would violate this assumption. Additional 
physical information is necessary to determine what happens at the event 
horizon, and the correct matching of interior to exterior geometry depends on
this physics, which may  or may not be consistent with (complex) analytic continuation 
of coordinates around the singularity at $r=r_{\!_M}$. 

The static Schwarzschild solution of an isolated uncharged mass was generalized 
to include electric charge by Reissner \& Nordstr\"om \cite{RN}, and more interestingly for 
astrophysically realistic collapsed stars, to include rotation and angular momentum
by Kerr \cite{Kerr}. The complete analytic extensions of the Reissner-Nordstr\"om and 
Kerr solutions were found as well \cite{BoyLinCar}. The global properties of these 
analytic extensions are more complicated and arguably even more unphysical than in the 
Schwarzschild case. For slowly rotating black holes with angular momentum $J < GM^2/c$, 
there are an {\it infinite} number of black hole interior and asymptotically flat exterior 
regions, singularities, and {\it closed timelike curves} in the interior region(s), which violate causality 
on macroscopic distance scales \cite{HawEll}. Again these apparently unphysical features 
appear in GR only if the mathematical hypothesis of complex analytic 
extension and continuation through real coordinate singularities is assumed.
Based on the Principle of Equivalence between gravitational and inertial mass, Einstein's 
theory possesses general coordinate invariance under all {\it regular} and {\it real} 
transformations of coordinates. It is the appending to classical General Relativity 
of the much stronger mathematical {\it hypothesis} of {\it complex} analytic 
continuation through {\it singular} coordinate transformations that leads to the 
global aspects of the Schwarzschild solution which may be unrealized in Nature.
This analytic continuation is generally invalid if there are stress-tensor sources 
encountered at or before the breakdown of coordinates.

In the general black hole solution characterized by mass $M$, angular 
momentum $J$, and electric charge $Q$, one can define a quantity called the 
{\it irreducible mass} $M_{irr}$ by the relation \cite{Mirr}
\be
M^2 = \left(M_{irr} + \frac{Q^2}{4G M_{irr}}\right)^2 + \frac{c^2 J^2}{4 G^2 M_{irr}^2}\,,
\label{MQJ}
\ee
The differential form of (\ref{MQJ}) is \cite{Smarr,BCH}
\be
dE = dMc^2=  \frac{c^2}{8\pi G}\,\kappa \,dA + \Omega\,  dJ + \Phi\,dQ
\label{diffSm}
\ee
which is the Smarr formula for a Kerr-Newman rotating, electrically charged black hole, 
in which 
\bes
\bea
&&\kappa = \frac{1}{M}\left[\frac{c^4}{4G} - \frac{4\pi^2 G}{c^4A^2} 
\left(Q^4 + 4c^2J^2\right)\right],\\
&&\Omega = \frac{4\pi J}{MA}\,,\\
&&\Phi = \frac{Q}{M} \left[\frac{c^2}{2G} + \frac{2\pi Q^2}{Ac^2}\right]\,,\\
&&A = \frac{16 \pi G^2}{c^4} M^2_{irr} =
\frac{4\pi G}{c^4}\,\left[2GM^2 -Q^2 + 2\sqrt{G^2M^4 - GM^2Q^2  - c^2J^2}\right],
\label{area}
\eea
\label{kOP}\ees
are the horizon surface gravity, angular velocity, electrostatic potential and area 
respectively. All dimensionful constants have been retained to emphasize that 
(\ref{diffSm})-(\ref{kOP}) are formulae derived from {\it classical} GR in which no 
$\hbar$ whatsoever appears. Notice also that the coefficient of $dA$ in (\ref{diffSm})
\be
\left(\frac{\partial E}{\partial A}\right)_{\!J, Q} = \frac{c^2\kappa}{8\pi G}
\ee
has both the form and dimensions of a {\it surface tension}.

\section{Quantum Black Holes and Thermodynamic Paradoxes}
\vspace{3mm}

It has been shown that in any classical process the irreducible mass $M_{irr}$ in (\ref{MQJ}), and 
therefore from (\ref{area}) the geometric area $A$ of the horizon can never decrease \cite{Mirr,HawA}. 
Since matter falling into a black hole would take its entropy with it, leading to an apparent violation
of the Second Law of Thermodynamics, but classically $A$ always increases, Bekenstein proposed 
that the black hole should itself be assigned an entropy proportional to $A$ \cite{Bek}.
Since $A$ does not have the units of entropy, it is necessary to divide the area by another 
quantity with units of length squared before multiplying by Boltzmann's constant $k_{_B}$, 
to obtain an entropy. However, classical GR (without a cosmological term) contains 
no such quantity, $G/c^2$ being simply a conversion factor between mass and 
distance. Hence Bekenstein found it necessary for purely dimensional reasons to 
introduce Planck's constant $\hbar$ and the Planck length 
$L_{Pl} = (\hbar G /c^3)^{\frac{1}{2}} = 1.616 \times 10^{-33}$ cm into the discussion,
proposing that the entropy of a black hole should be
\be
S_{_{BH}} = \gamma k_{_B} \frac{A}{L_{Pl}^2} = \gamma \,\frac{16 \pi G k_{_B}}{\hbar c} \, M_{irr}^2\,,
\label{SBHg}
\ee
with $\gamma$ a constant of order unity \cite{Bek}. He showed that {\it if} such 
an entropy were assigned to a black hole, so that it is added to the entropy of
matter, $S_{tot} = S_m + S_{_{BH}} $,  then this total generalized entropy would
plausibly always increase. In fact, this is not difficult at all, and the generalized
second law $\Delta S_{tot} \ge 0$ is usually satisfied by a very wide margin, simply
because the Planck length $L_{Pl}$ is so tiny, and the macroscopic area of a black hole 
measured in Planck units is so enormous. Hence even the small increase of mass 
and area caused by dropping into the black hole a modest amount of matter 
and concomitant loss of matter entropy $\Delta S_m < 0$ is easily overwhelmed 
by a great increase in $S_{_{BH}}$, $\Delta S_{_{BH}} \gg |\Delta S_m|$, 
guaranteeing that the generalized total entropy increases: $\Delta S_{tot} > 0$.

Soon after Bekenstein's proposal Hawking argued that black holes would also
emit radiation at a characteristic temperature \cite{HawT}
\be
T_{_H} = \frac{\hbar\kappa}{2\pi c k_{_B}}\  \,\stackrel{J=Q=0}{=}\ \,
 \frac{\hbar c^3}{8 \pi G k_{_B} M}\,,
\label{TH}
\ee
where the first equality is general, and the second equality applies only 
for a Schwarzschild black hole with $J=Q=0$. With the temperature inversely 
proportional to its mass assigned to a black hole by this formula, if 
we interpret (\ref{diffSm}) as the first law of thermodynamics in the form
\be
dE = dM c^2 = T_{_H} \, dS_{_{BH}} + \Omega\,  dJ + \Phi\,dQ\,.
\label{dETdS}
\ee
then the coefficient $\gamma$ in (\ref{SBHg}) is fixed to be $1/4$. Notice that this
involves simply multiplying $\kappa$ and dividing $A$ in the {\it classical}
Smarr formula (\ref{diffSm}) by $\hbar$. 

Since the temperature $T_{_H} < 0.1\mu K$ for a solar mass black hole is so exceedingly small, 
the prospects for testing the Hawking prediction directly by observation are correspondingly remote. 
Nevertheless the formula (\ref{dETdS}) is simple and appealing, and it has been generally 
accepted since soon after Hawking's paper appeared. However, simultaneously and from the 
very beginning, a number of problems with this thermodynamic interpretation made their 
appearance as well. 

Firsty, the entropy (\ref{SBHg}) is non-extensive both in not being proportional to the volume
but to the area, and in being entirely independent of the number of particle species. This
points to a basic problem in trying to account for the the entropy by a microcanonical
counting of microstates according to Boltzmann's formula, 
\be
S = k_{_B} \ln W(E)\,.
\label{Boltz}
\ee
Normally one would expect the number of distinct microscopic states $W(E)$ at a given total energy $E$, 
and hence the entropy, to depend on the number of distinct particle species.

Secondly, tracing the thermal quanta at asymptotic temperature $T_{_H}$ backwards 
indicates that at late times they must have originated very close to the event horizon at $r=r_{\!_M}$ with local 
energy
\be
\hbar \,\omega_{loc} \sim \frac{k_{_B} T_{_H}}{f^{\frac{1}{2}}(r)} 
= \frac{\ \ k_{_B} T_{_H}}{\sqrt{1 - \frac{r_{\!_M}}{r}}} \ \rightarrow \ \infty\,,
\label{locfreq}
\ee
which is arbitrarily large, and in particular exceeds the Planck energy $\hbar c/L_{Pl}$, at which 
the semi-classical approximation of a fixed classical background geometry would be expected to fail.

Thirdly, as pointed out by Hawking himself \cite{Hawcv}, a temperature inversely 
proportional to the mass $M = E/c^2$ implies that the heat capacity of a Schwarzschild  black hole,
\be
\frac{dE\ }{dT_{_H}} = - \frac{8 \pi G k_{_B} M^2}{\hbar c} = - \frac{Mc^2}{T_{_H}} < 0
\label{dEdT}
\ee
is {\it negative}, whereas in statistical mechanics the heat capacity of any system 
in stable equilibrium is related to the energy fluctuations about its mean value
$\lag E \rag$ by (at constant volume)
\be
c_{_V} = \left(\frac{d\lag E\rag}{dT}\right)_{_V} = 
\frac{1}{k_{_B}T^2} \,\Big\lag (E- \lag E\rag)^2\Big\rag > 0\,,
\label{cV}
\ee
and is necessarily positive \cite{Maz}. A negative heat capacity can be formally obtained in the
microcanonical treatment of certain non-relativistic gravitational systems such as globular 
clusters with very long relaxation times that have not yet reached a true equilibrium state \cite{LB,Pad}.
However whereas the relaxation time scale for a globular cluster due to two-body stellar encounters 
can be much larger than the dynamical time scale $\sqrt{R^3/GM}$ (with $R$ a typical dimension of 
the cluster and $M$ its mass), statistical fluctuations in the Hawking thermal flux occur on the 
dynamical time scale $2GM/c^3$ itself,  which is very short ($\sim 10^{-5}$ sec for a solar mass 
black hole). This is indeed the time scale for an unstable mode around the equilibrium state
to develop \cite{BHnegmode}. It is therefore by no means clear that equilibrium thermodynamics can 
be applied to black holes at all. 

Finally, a cold quantum system generally has a low entropy since thermal excitations are 
suppressed at zero temperature. However in the limit $M \rightarrow \infty$ or $\hbar \rightarrow 0$, 
for which (\ref{TH}) gives $T_{_H} \rightarrow 0$, (\ref{SBHg}) gives $S_{_{BH}} \rightarrow \infty$,
{\it i.e.} an {\it infinitely large} entropy at {\it absolute zero} temperature!

It is instructive to evaluate $S_{_{BH}}$ for typical astrophysical black holes. Taking as our unit of mass 
the mass of the sun, $M_{\odot} \simeq 2 \times 10^{33}$ gm., we have
\be
S_{_{BH}} \simeq 1.050 \times 10^{77}\, k_{_B}\, \left(\frac{M}{M_{\odot}}\right)^2\,.
\label{SBH}
\ee
This is truly an enormous entropy. For comparison, we may estimate the entropy of the sun as it is, 
a hydrogen burning main sequence star, whose entropy is given to good accuracy by the entropy 
of a non-relativistic perfect fluid. This is of the order $N k_{_B}$ where $N$ is the number of nucleons 
in the sun $N \sim M_{\odot}/m_{_N} \sim 10^{57}$, times a logarithmic function of the density and 
temperature profile which may be estimated to be of the order of $20$ for the sun. Hence the entropy 
of the sun is roughly
\be
S_{\odot} \sim 2 \times 10^{58} \, k_{_B}\,,
\label{Ssun}
\ee
or nearly $19$ orders of magnitude smaller than (\ref{SBH}).  Since (\ref{SBH}) makes no
reference to how a black hole is formed, we may in principle imagine forming one from the sun
adiabatically by an arbitrarily slow contraction. At every time in this adiabatic process the
entropy of the sun would remain (\ref{Ssun}). At the instant that the event horizon is reached
this entropy would have to jump discontinuously somehow to (\ref{SBH}). From Boltzmann's formula 
(\ref{Boltz}) this means that the number of microstates of a black hole must jump by $\exp(10^{19})$ 
at that instant at which the event horizon is reached, a truly staggering proposition. The loss of the 
information represented by this enormous jump in entropy is one form of the ``information paradox," 
which is so serious that it led Hawking to suggest that quantum mechanics itself must be 
altered \cite{Hawunit}. This is all the more strange when one considers that according to classical 
GR by a change to Kruskal-Szekeres coordinates (\ref{SKSz}) nothing at all is supposed to happen 
at the event horizon.

For all of these reasons the thermodynamic interpretation of (\ref{dETdS}) remains 
problematic. On the other hand, if the inserted $\hbar$ in (\ref{dETdS}) is simply cancelled 
and one returns to the differential Smarr relation (\ref{diffSm}), derived from classical GR, 
these difficulties immediately vanish. One would only be left to explain in what sense
the surface gravity $\kappa$ is a surface tension. 

If collapse to a black hole does not produce an equilibrium state, because of its negative heat 
capacity and rapidly growing unstable negative mode, the question then naturally arises 
of what is the final equilibrium state of complete gravitational collapse? In condensed
matter physics at high density and low temperatures quantum effects always play an important 
role, and lead generally to a phase transition in which a macroscopic condensate is formed. 
In the proposal of a quantum phase transition producing a gravitational Bose-Einstein 
condensate (GBEC) in the interior of a fully collapsed object a physical surface and surface 
tension replaces the mathematical horizon of classical black hole, and removes all of the 
thermodynamic paradoxes of quantum black holes \cite{PNAS}.

\section{Quantum Effects at Horizons: The Role of the Trace Anomaly}
\vspace{3mm}

The analytic continuation of the static Schwarzschild geometry to the black hole interior by
Kruskal coordinates (\ref{SKSz}) assumes the complete absence of sources to Einstein's eqs. in the
vicinity of the horizon. However, it has long been known that at least in certain states quantum stresses
can be quite important there. In the natural ``vacuum" state empty of all particles at infinity and corresponding 
to the static Schwarzschild time $t$, called the Boulware state $\vert B\rag$ \cite{Boul}, the expectation 
value of the renormalized stress tensor \cite{ChrFul}
\be
\langle B \vert T^a_{\ b} \vert B \rangle_{_{R}} \rightarrow - \frac{\pi^2}{90} 
\frac{\hbar c}{(4 \pi r_{\!_M})^4}
\, \left(1 - \frac{r_{\!_M}}{r}\right)^{-2}\, {\rm diag} \, (-3, 1, 1,1)\,,
\label{TBoul}
\ee
becomes arbitrarily large as $r \rightarrow r_{\!_M}$. Being a dimension four operator,
the $\left(1 - \frac{r_{\!_M}}{r}\right)^{-2}$ behavior of $\lag B \vert T^a_{\ b} \vert B \rag_{_{R}}$ 
is a kinematic consequence of the blueshift (\ref{locfreq}). Clearly, if such a state were even close
to being realized in practice, its stress-energy would act as a significant physical source for the 
semi-classical Einstein's equations,
\be 
R^a_{\ b} - \frac{R}{2}\, \delta^a_{\ b}  + \Lambda \delta^a_{\ b} = 8\pi G\, \lag T^a_{\ b}\rag_{_R}\,.
\label{scE}
\ee
and necessarily influence the background spacetime (\ref{sphsta}). The assumption that
$T^a_{\ \,b} = 0$ on the horizon upon which the analytic continuation of coordinates in (\ref{SKSz})
critically depends would have to be re-evaluated, or discarded entirely, leading potentially
to a very different interior solution.

Let us dispel at this point any concern that large quantum effects on the horizon violate coordinate 
invariance or the Equivalence Principle. It is only some classical notions about locality that are 
violated by such effects and in the same way that quantum mechanics itself usually violates them. 
The quantum effective action in a background field, be it electromagnetic or gravitational,
is generally a non-local functional of the background. It is this non-locality which makes particle
creation in electromagnetic fields or in curved spacetimes possible. For this same reason
the stress tensor (\ref{TBoul}) derived from the one-loop quantum effective action 
depends on boundary conditions in the full spacetime, not simply local quantities
such as the Riemann curvature (\ref{RieS}), which are small on the horizon. Because of
quantum mechanics matter behaves as waves and it is these virtual matter waves in 
the macroscopic quantum state, non-local and coherent on the horizon scale that lead 
to (\ref{TBoul}). In a {\it different} (Unruh) state it is exactly the same non-local wavelike 
nature of matter that leads to the Hawking particle creation process.

If there were but a single ``vacuum" state of matter in a Schwarzschild background that showed
evidence of large quantum effects near the horizon, it might be possible to argue that
the single Boulware state $\vert B\rag$ should be excluded as pathological. In fact, just 
the opposite is true. By considering the most important non-local terms in the quantum effective
action, related to the conformal anomaly, it is possible to show that {\it most} states
behave as in (\ref{TBoul}) \cite{MotVau,Zak}. 

That the important quantum effects near horizons are associated with the conformal or 
trace anomaly is related to the fact that due to the kinematic blueshift (\ref{locfreq})
becoming greater than all other mass scales as the horizon is approached,
all quantum fields, massless or not, become essentially conformal in the horizon region.
Quantitatively, it has recently become possible to calculate these effects from the
form of the quantum effective action due to the anomaly,
\bea
&&S_{anom}=\frac{b'}{2}\int d^4x\sqrt{-g}\left\{
-\left(\sq \varphi\right)^2 + 2\left(R^{ab} - \frac{R}{3} g^{ab}\right)
\nabla_a \varphi\nabla_b\varphi + \left(E - \frac{2}{3} \sq R\right)
\varphi\right\}   \label{SEF}  \\
&&\hspace{-6mm}+  b\int d^4x\sqrt{-g}\left\{\! -\!\left(\sq \varphi\right)\!
\left(\sq \psi\right) + 2\left(R^{ab} - \frac{R}{3}g^{ab}\right)\!\nabla_a \varphi
\nabla_b \psi + \frac{1}{2} C_{abcd}C^{abcd}
\varphi + \frac{1}{2}\! \left(E - \frac{2}{3} \sq R\right)\! \psi \right\}\nonumber
\eea
in terms of curvature tensors and two new scalar field degrees of freedom $\varphi$ and 
$\psi$ \cite{MotVau}.The form of this effective action is determined by general principles 
of covariance and the form of the trace anomaly, with only the coefficients $b$ and $b'$ 
dependent upon the number and spin of the underlying quantum fields \cite{Duff}. The 
action (\ref{SEF}) is clearly a spacetime scalar and therefore in that precise sense it is 
certainly consistent with the Equivalence Principle. 

If (\ref{SEF}) is varied with respect to the scalar fields $\varphi$ and $\psi$, one 
obtains their (linear) eqs. of motion. Solving these linear differential
eqs. formally in terms of the Green's functions of the relevant differential operator
and inserting this formal solution for $\varphi$ and $\psi$ back into (\ref{SEF}) allows 
one to rewrite the effective action in its original non-local but still covariant form.
This non-local effective action of the trace anomaly in four dimensions is the analog 
of the two-dimensional anomaly effective action \cite{Poly} which can also be put into
 a local form with the addition of a new local scalar degree of freedom \cite{NJP}.

The eqs. for the scalar fields $\varphi (r)$ and $\psi(r)$ derived from (\ref{SEF}) 
can also be solved in the static spherically symmetric Schwarzschild background,
and the results substituted into the the stress tensor also obtained from (\ref{SEF})
by varying with respect to the metric. This stress tensor incorporating the one-loop
quantum effects of the anomaly generally exhibits divergent behavior at the horizon, 
just as in (\ref{TBoul}), for a wide variety of states \cite{MotVau}. This verifies explictly
the origin of such divergent behavior as the one-loop non-local quantum effective
action of the anomaly, and the generic behavior of the divergences of the stress tensor 
near $r=r_{\!_M}$ due to quantum effects. 

Let us remark also that it is by no means 
necessary for the state to have truly divergent behavior on the horizon as in (\ref{TBoul}) 
in order to have large quantum effects there. Any state in which $G\,  \lag T^a_{\ b}\rag_{_R}$ 
becomes of order of the (small) classical curvature (\ref{RieS}) near the horizon is already 
enough to produce significant backreaction effects on the geometry. To see this, note that 
Einstein's equations (\ref{scE}) in the general static spherically symmetric metric (\ref{sphsta}) 
imply
\be
\frac{1}{r}\frac{d}{dr} \left(\frac{h}{f}\right) = - 8 \pi G \ \frac{\rho + p}{f}\,,
\label{hfeq}
\ee
while the Riemann curvature tensor has the component
\be
R^{tr}_{\ \ tr} = \frac{h}{4} \left( \frac{f^{\prime 2}}{f^2} - \frac{2f''}{f} - 
\frac{h'}{h}\frac{f'}{f}\right)
\label{Riem}
\ee
(where primes denote differentiation with respect to $r$). In a pure vacuum, $h(r) = f(r)$
and (\ref{Riem}) becomes $-f''/2$ and hence remains finite. But from (\ref{hfeq})
if $\rho + p > 0$ in the horizon region where $f$ or $h$ vanishes (or nearly vanishes), 
in general  $h \neq f$ and the cancellation of the singularities in (\ref{Riem}) will not occur.
In that case even the Riemann curvature invariant (\ref{RieS}) can become large. The 
horizon region defined by one or more metric functions $h$ or $f$ approaching zero is
extremely sensitive to matter source perturbations, and even ``small" quantum vacuum 
polarization effects can have relatively large effects on the spacetime geometry in this region.
The anomaly scalar fields $\varphi(r)$ or $\psi(r)$ behaving like $\ln f(r) \rightarrow \infty$
shows that non-local quantum effects can easily produce significant perturbations near
the horizon.

In flat space one can understand the large quantum effects of the anomaly on the light cone
(hence a null surface such as an event horizon of a black hole) as the result of the massless 
poles of the conformal $\varphi$ and $\psi$ scalar propagators, which describe correlated 
two-particle states of the underlying quantum fields at threshold \cite{GiaMot}.  Two-particle 
pairing at the Fermi surface is common in many-body physics, where it is responsible for 
superfluidity and superconductivity at low temperatures. In these non-relativistic quantum 
systems as the temperature is lowered or the pressure raised there is a phase transition to a 
condensate, in which a macroscopically large number of paired fermions or bosons collapse 
into a single coherent quantum state. In relativistic quantum field theory, anomalies (both axial 
and conformal) seem to be the only cases where two-particle pairing to form a spin zero bosonic 
state at the kinematic threshold $k^2 = 0$ of the Dirac sea (for a fermion pair) can occur consistent 
with Lorentz or general coordinate invariance. In this case, of course, it is the vacuum itself which 
becomes unstable to condensation. The strong effects of these anomalous quantum two-particle 
correlations near black hole horizons invites us to consider the possibility that a phase transition 
in spacetime itself is possible, in which the black hole interior is replaced by a kind of low temperature 
gravitational vacuum condensate. 

\section{Gravitational Vacuum Condensate Stars}
\vspace{3mm}

The region near the horizon $r=r_{\!_M}$ in which quantum effects can be important is very narrow.
The spatial extent of the region can be estimated by asking how close to $r_{\!_M}$ must $r$ be
for the quantum stresses (\ref{TBoul}) to give curvature corrections through Einstein's eqs. (\ref{scE})
of the same order as the classical Riemann curvature (\ref{RieS}). This gives the estimate
\be
\vert f(r_{\!_M} \pm \delta r)\vert  = \Big\vert 1 - \frac{r_{\!_M}}{r_{\!_M} \pm \delta r} \Big\vert
\simeq  \frac{\delta r}{r_{\!_M}}
\equiv \epsilon \simeq \frac{M_{Pl}}{M} \simeq 10^{-38}\,\frac{M_\odot}{M}\,,
\ee
in solar mass units. Thus $\delta r \simeq L_{Pl}$. However because the {\it physical} distance is given 
by the line element (\ref{sphsta}) this corresponds to a physical distance
\be
\ell \simeq \frac{\delta r}{f^{\frac{1}{2}}} \simeq \sqrt{L_{Pl}\, r_{\!_M}} \simeq 
3\,\times\,10^{-14}\, {\sqrt\frac{M}{M_\odot}} \ {\rm cm}\,,
\label{elldef}
\ee
of the order of the diameter of an atomic nucleus for $M$ of a few solar masses, which is tiny on
astrophysical scales but still much larger than the microscopic Planck length. Well away from this 
boundary layer $f(r) = (1- r_{\!_M}/r) \rightarrow 1$, and quantum effects are of order $\epsilon^2$, 
and hence completely negligible. Thus we expect the general conditions for an effective field
theory treatment to hold everywhere except within this thin boundary layer of thickness $\ell$, where
the quantum effects are relatively large. Mathematically, this situation is similar to those encountered
in hydrodynamic shock waves where flows can be accurately described by the continuum
Navier-Stokes eqs. except for a thin boundary layer inside the shock front of order of the mean
free path of the molecular constituents of the fluid where the continuum approximation breaks
down. A second example from fluid mechanics of a boundary layer is the Prandtl layer of the
flow of a fluid around an obstacle, such as a ship's bow or an airplane's wing, which is responsible
for the full macroscopic drag on the body. In both of these situations one can obtain an
accurate description in the continuum fluid description, provided one supplements it with higher 
derivative terms usually neglected and/or proper boundary conditions derived from conservation 
laws across the layer.

Away from the boundary layer on either side of it, Einstein's eqs. (\ref{scE}) apply with negligible
quantum effects. However, because the layer is a phase boundary, in which the quantum
vacuum itself has changed its character, the parameters of the low energy description in terms
of classical General Relativity need not have the same values on opposite sides of the phase
boundary. On general grounds the ``latent heat" of the ground state of a system can change
in a phase transition, and for the vacuum itself this is the cosmological term $\Lambda$ in
Einstein's eqs. (\ref{scE}). Moreover it has been shown that the fluctuations of the anomaly
scalar degrees of freedom in (\ref{SEF}) which are responsible for the large quantum effects
near the horizon also give rise to the cosmological term becoming {\it dynamical}, varying in
space and time \cite{NJP,AntMot}. A non-zero $\Lambda$ can also be interpreted 
thermodynamically as a zero entropy density condensate because of the Gibbs' relation
\be
\rho + p = sT + \mu n\,.
\label{Gibbs}
\ee
Since $\Lambda > 0$ corresponds to a positive vacuum energy density with negative pressure
$p = - \rho$, the entropy density $s$ must vanish if there is no conserved number density or
associated chemical potential $\mu = 0$. 

We are familiar with phenomenological low energy
effective field theories such as the electroweak theory of the Standard Model where the vacuum
energy and effective $\Lambda$ changes due to the Higgs field developing a non-zero condensate 
$\lag \Phi \rag \neq 0$ which is a pure quantum state with zero entropy, so that $W =1$ and $S=0$
in (\ref{Boltz}). In the bag model of hadrons there is a vacuum energy, the bag constant, associated 
with the rapid crossover (not a true phase transition) between the interior of a hadron where quarks 
and gluons are approximately free and the exterior where they cannot propagate. Note also that
in all these cases the vacuum condensate eq. of state $p= -\rho$ with $\rho > 0$ {\it violates} the strong 
energy condition $\rho + 3 p \ge 0$ which is used to prove the classical black hole singularity 
theorems \cite{HawEll}.

Thus it is reasonable to assume that the effective value of $\Lambda$ will change at the quantum
phase transition and become non-zero in the interior of the quantum boundary layer at $r \simeq r_{\!_M}$.
Einstein's eqs. again apply within with this new value of $\Lambda$ describing the interior vacuum 
condensate state. Just as the vacuum Einstein's eqs. possess a static, spherically symmetric solution 
for an isolated mass, namely the Schwarzschild solution (\ref{Sch}), they possess a static, spherically 
symmetric vacuum solution for positive cosmological term, namely the de Sitter solution, (\ref{sphsta}) 
with \cite{deS}
\be
f(r) = h(r) = 1 - \frac{r^2}{r_{\!_H}^2}\,, \qquad{\rm with} \qquad r_{\!_H} \equiv \frac{1}{H} 
= \sqrt{\frac{3}{\Lambda}}\,.
\label{deS}
\ee
Just as the exterior Schwarzschild geometry will attract all classical matter into its horizon at $r=r_{\!_M}$, 
the de Sitter geometry will sweep out all classical matter outwards towards its horizon at $r=r_{\!_H}$.
Thus a strictly static solution of Einstein's eqs. is possible if both the exterior and condensate interior
are free of other matter or stress energy sources. As the de Sitter horizon is approached from the 
interior there are again quantum effects of the anomaly action and stress tensor in generic
quantum states that grow like $f^{-2}(r) = (1-H^2r^2)^{-2}$ \cite{MotVau,DSAnom}. Hence a 
globally static solution is possible only if the interior and exterior geometries are matched in
their mutual near horizon regions: $r_{\!_H} \simeq r_{\!_M}$. This fixes the interior vacuum 
condensate energy density to 
\be
\rho_{cond} = \frac{3c^4 H^2}{8 \pi G} = \frac{3 \,c^8}{32\pi G^3M^2}\,,
\label{rhocond}
\ee
so that the total mass of the interior condensate is
\be
\frac{4\pi r_{\!_H}^3}{3}\,\frac{\rho_{cond}}{c^2} = M \,,
\ee
the total mass of the Schwarzschild solution in the exterior. This static matching of the interior
de Sitter vacuum condensate (\ref{deS}) to the exterior Schwarzschild geometry (\ref{Sch})
is what we have called a gravitational vacuum condensate star or {\it gravastar} \cite{PNAS,CHLS}.

A key point about matching an interior de Sitter solution to an exterior Schwarzschild solution
is that this matching cannot be achieved across a null surface such as an horizon because 
such a null surface can contain radiation leading to arbitrary discontinuities
of the metric functions $f(r)$ and $h(r)$. However, once the null horizon singularity of these
functions is regulated by a {\it finite} boundary layer of order $\ell$ in (\ref{elldef}), then the
entire boundary layer is a {\it timelike} tube. In ref. \cite{PNAS} this was done by
application of the Israel junction conditions, assuming a $p=\rho$ relativistic
fluid eq. of state within the boundary layer  \cite{Isr}. A more accurate matching is now
possible with the effective action (\ref{SEF}) and stress tensor derived from it.
The stability question can then be re-examined through the Lagrangian variational
principle the effective action provides.

The gravastar proposal for the final equilibrium quantum ground state for complete gravitational
collapse eliminates all of the paradoxes associated with black holes. Since the interior
is described by the static patch of de Sitter spacetime (\ref{deS}), there is evidently no singularity
and no place for information to disappear into. Quantum unitarity is preserved. Since
the interior is a pure quantum vacuum state which can be regarded as a macroscopic
condensate, it has zero entropy. Any entropy is associated with the fluctuations present
in the quantum boundary layer at $r \simeq r_{\!_M}=r_{\!_H}$, and may be estimated to
be of order  \cite{PNAS}
\be
S \sim \frac{\rho_{cond}}{T_{_H}}\, \ell A \sim k_{_B} \left(\frac{M}{M_{Pl}}\right)^{\frac{3}{2}} \simeq 
10^{57}\, k_{_B} \left(\frac{M}{M_{\odot}}\right)^{\frac{3}{2}}\,,
\ee
which is within an order of magnitude or so of a typical stellar progenitor. The
$M^{\frac{3}{2}}$ dependence on mass is consistent with estimates from ordinary
thermodynamics for a relativistic star (see e.g. \cite{Zel}). Hence there is no enormous 
mismatch of entropies to be explained as for (\ref{SBH}), and no information paradox.
There is also no Hawking radiation from a gravastar, the black hole event horizon
having been replaced with a boundary layer or thin shell of finite thickness (\ref{elldef})
and finite energy density. Although this energy density is enormous by terrestrial
standards, it is of the same order as that of a neutron star, {\it i.e.} a bit greater than
nuclear energy density, but far below Planck scale energy densities. Hence there is no 
trans-Planckian problem and the effective field theory of Einstein's equations
(\ref{scE}) augmented by the quantum effects of the anomaly action (\ref{SEF}) 
should be sufficient to describe the main macroscopic features of gravastars.
The final state of a completely cooled gravastar as its temperature approaches
absolute zero is a quantum stable ground state of zero entropy similar to
a finite sized Bose-Einstein condensate of cold atoms at absolute zero, trapped
by its own self-gravity.

As astrophysical objects gravastars would be cold, dark and compact, and therefore
mimic classical black holes in almost every respect observationally. In complete
isolation it would be virtually impossible to tell a gravastar from a black hole.
The accretion of matter onto a gravastar would be similar to that of a black hole
except at the very last stage when it interacts with the boundary layer. At the
boundary layer, matter can be converted into pure energy and baryon and 
lepton number is likely also violated by anomalous processes in the Standard
Model. Extensions to more realistic solutions with rotation, magnetic fields and 
taking into account Standard Model matter interactions through the anomaly effective 
action (\ref{SEF}) are possible. Although the release of energy by
matter striking the boundary layer can be very large, since the boundary layer 
thickness (\ref{elldef}) is so small, any radiation from a {\it non-rotating} gravastar
would be extremely redshifted out of $\gamma$-rays or X-rays to much lower energies. 
Thus the standard arguments for the absence of a surface \cite{Nar} do not apply \cite{Abr}. 
Gravitational waves should be generated at the natural oscillation frequencies of
the physical thin shell, and perhaps would provide the cleanest evidence of
the surface distinguishing a gravastar from a black hole. Once rotation, particularly 
rapid rotation is considered, the situation may be quite different, and the efficient 
conversion of matter into energy could conceivably provide the central powerhouse
for Gamma Ray Bursters and some of the most energetic sources observed in
the universe. 

\section{A New Model for Cosmological Vacuum Energy}
\vspace{3mm}

The scalar $\varphi$ and $\psi$ fields couple to the metric particularly strongly on the cosmological 
Hubble horizon scale where they have their largest effect. Thus the fluctuations of the scalar 
degrees of freedom determined by the anomaly may lead to a phase transition to precisely the 
conformally invariant phase of gravity described by the fixed point $\Lambda = 0$ \cite{NJP,AntMot}
in the near vicinity of the horizon. Since the interior condensate phase has the eq. of state $p = -\rho$, 
exactly that of the cosmological dark energy observations suggest pervade our universe \cite{SN},
and the gravastar solution automatically fixes the vacuum condensate energy density 
$\rho_{cond}$ in terms of its size by (\ref{rhocond}), an interesting possibility raised by this 
solution is that the observable universe itself could be the interior of a gravastar. In that case 
the observed cosmological dark energy of our universe, some $72\%$ of the critical energy 
density $\rho_{crit}$ defined by the present value of the Hubble parameter $H_0$ would be 
identified with the condensate energy density 
\be
\rho_{cond} \simeq (0.72)\, \rho_{crit} = (0.72)\, \frac{3c^4 H_0^2}{8\pi G} 
\simeq 6.5 \times 10^{-9} \ {\rm erg/cm}^3\,.
\label{cond}
\ee
If our location is far from the bubble wall, it would be difficult to distinguish the cosmological model 
of a spherical bubble of vacuum energy condensate from the standard Friedmann-Robertson-Walker 
(FRW) model, although on the cosmological Hubble horizon scale $H_0^{-1} \simeq 4.2$ Gpc and
of course globally it would be dramatically different, with a preferred origin and a physical surface. 
At the horizon the quantum fluctuations of the scalar degrees of freedom described by the anomaly 
action (\ref{SEF}) would be large. In ref. \cite{DSAnom} it was pointed out that these anomaly
scalars couple to the metric near the horizon and generate fluctuations that are similar to
those in inflationary models. Thus it may be that these fluctuations emanating from the horizon
are what we observe in the Cosmic Microwave Background (CMB) anisotropies, without any
need to introduce an inflaton. Moreover, since the fluctuations of the anomaly scalars lead to a 
conformally invariant phase of gravity \cite{NJP,AntMot}, the fluctuations should have a spectrum 
and statistics consistent with conformal invariance, including in their non-Gaussian features. 
Thus, one way of testing this cosmological model would be to detect the non-Gaussian bi-spectrum 
of the CMB with the shape predicted by conformal invariance \cite{sky,Mor,CMB}.

The new cosmological model is then a kind of bubble or `bag' of vacuum energy, containing these
gravitationally coupled degrees of freedom, which is well described as approximately 
a de Sitter universe of finite size of the order of the Hubble radius. Like an ordinary bubble 
of gas, the pressure inside is determined by the size of the bubble and its surface tension, 
not any UV cutoff, and the interior pressure adjusts itself dynamically to the boundary 
conditions at the Hubble scale. In the effective field theory of Einstein gravity augmented
by the anomaly action (\ref{SEF}), the vacuum energy of infinite flat space is identically zero, 
and the residual small $\rho_{cond}$ of (\ref{cond}) is a boundary effect of a macroscopic 
condensate. Then the essential cosmological problem becomes to explain why $\Omega_{cond}  = 
\rho_{cond}/\rho_{crit} \simeq 0.72$ rather than unity, and how the vacuum energy and pressure 
adjusts itself dynamically at the boundary to maintain $\Omega_{cond} \approx 1$  over time. 

Clearly additional details of the gravastar model incorporating matter and radiation need to be 
worked out and it then must pass the many successful tests of the usual FRW big bang model 
before it can be considered a realistic alternative cosmology. The essentially automatic
incorporation of dark energy with the approximately correct value and solution of the
cosmological dark energy problem would seem to make the model worth pursuing further.

\ack

I gratefully acknowledge all my collaborators on whose work this talk was based, and 
especially P. O. Mazur, with whom the hypothesis of gravitational vacuum condensate stars 
was developed. I also thank Jos\'e Luis Jaramillo, Carlos Barcel\'o, and all the organizers
of the Spanish Relativity Meeting ERE2010 at which this talk was presented.

\end{document}